\documentclass[reprint,superscriptaddress,aps,prb,showpacs,floatfix]{revtex4-1}
\usepackage{graphicx}%
\usepackage{dcolumn}%
\usepackage{bm}
\date{\today}
\begin{document}
%\linenumbers
\title{Superconducting tantalum nitride-based normal metal-insulator-superconductor tunnel junctions}
\author{S. Chaudhuri}  %\email{sachaudh@jyu.fi }
\author{I. J. Maasilta} \email{maasilta@jyu.fi }
 \affiliation{Nanoscience Center, Department of Physics, University of Jyv\"askyl\"a, P. O. Box 35, FIN-40014 University of Jyv\"askyl\"a, Finland}
\begin{abstract}
We report  the development of  superconducting tantalum nitride (TaN$ _{x} $)  normal metal-insulator-superconductor (NIS) tunnel junctions. For the insulating barrier, we used both AlO$ _{x}$ and TaO$ _{x}$ (Cu-AlO$ _{x}$-Al-TaN$ _{x} $ and Cu-TaO$ _{x}$-TaN$ _{x} $), with both devices exhibiting temperature dependent current-voltage  characteristics which  follow the simple one-particle  tunneling model. The superconducting gap follows a BCS type temperature dependence, rendering these devices suitable for sensitive thermometry and bolometry from the  superconducting transition temperature $T_{\text{C}}$ of the TaN$ _{x} $ film at $\sim 5$ K down to  $\sim$ 0.5 K. Numerical simulations were also performed to predict how junction parameters should be tuned to achieve electronic cooling at temperatures above 1 K. %The characteristic device parameters, namely the effective broadening of the   density of states (DOS) of the superconductor ($ \Gamma $) and  the specific tunneling resistance ($ r_S $), were extracted from theoretical fits.  In AlO$ _{x}$ barrier devices, the demonstrated $ \Gamma $ values were similar to those of  niobium (Nb) and niobium nitride (NbN$ _{x} $) based devices, while  $ r_S $ was about 2 orders of magnitude smaller than for the NbN$ _{x} $ counterparts fabricated in a similar manner. Numerical simulations predict that a further lowering of $ r_S $ would lead to a sizeable electronic cooling at temperatures around $0.4 T_C$ even for the observed broadening, and an additional further reduction of $\Gamma$ would strongly improve the performance. 
\end{abstract}
\maketitle
%=============================================INTRODUCTION===============================================

Normal metal-insulator-superconductor (NIS) tunnel junction devices  aimed at low temperature thermometry, bolometry and refrigeration have witnessed  significant developments in the last decade\cite{RMP,R2}. %,PanuPRL,Miller}.  
Aluminum (Al)  based NIS devices  already offer  sensitive thermometry in the sub 1 K range \cite{Nahum}, and significant cooling power approaching 1 nW at $ \sim $ 0.3 K has recently been demonstrated \cite{Lowellnew,Nguyen}. In addition to direct electronic cooling, sizeable indirect phonon cooling of  suspended membranes \cite{Luukanen, Clark, Miller}, beams \cite{PanuPRL},  and a general-purpose refrigeration platform \cite{Lowellnew} have been achieved using Al coolers.  However, to achieve operation at higher temperatures, one must switch to materials with higher superconducting transition temperatures ($ T_{\text{C}} $) than the $T_{\text{C}}$ of Al at $\sim 1.2$ K, as $T_{\text{C}}$ limits the maximum range of thermometry, and cooling power drops strongly above $T \sim 0.4 T_{\text{C}}$ \cite{RMP}. %While the cooling power of such a device is directly proportional to the square of the superconducting gap ($ \Delta $) and the thermometry range is determined by the superconducting transition temperature  ($ T_C $),  device parameters like  specific tunneling resistance ($r_S$) of the NIS junction and  the effective  broadening of the density of  states ($ \Gamma $) \cite{Dynes} of the superconductor profoundly  influences their cooling and thermometry performance.  %In order to exploit the expected improvement  in   cooling and thermometric performance for devices employing  intermediate $ T_C $ superconductors, 

Recently, we fabricated   Nb ($ T_{\text{C}} $ $ \sim $ 8 K) \citep{Nb} and NbN$ _{x} $ ($ T_{\text{C}} $ $ \sim $ 12 K) \citep{NbN} based NIS devices and demonstrated an order of magnitude increase of the thermometry range, and an observation of some electronic cooling in the Nb device\citep{Nb}. However, as the optimal operational temperature for cooling for those type of devices is aroung 3.5 - 5 K, one should also develop NIS devices with a $T_{\text{C}}$ in the intermediate range between Al and Nb/NbN devices. This is important because the cooling power also deteriorates fast when the operational temperature is lower than the optimal, and thus Nb or NbN based coolers may not be able to work effectively enough in the temperature range 1 - 3 K.  % The ideal $T_C$ for such a operational temperature is thus around 3 K. 
Here, we experimentally demonstrate that tantalum nitride (TaN) with a $T_{\text{C}} \sim $ 5K can be used as the superconducting electrode in a micron-scale NIS device. The thermometric characteristics were essentially ideal in the temperature range 0.5 - 5 K, and the observed specific tunneling resistance and the broadening of the superconducting density of states  were reasonably low, giving us hope of also developing electronic coolers in the temperature range 1 - 3 K in the future. This was elaborated by numerical simulations, which demonstrated that a further lowering of the specific tunneling resistance of the junctions (in principle a straightforward process) would lead to a sizeable electronic cooling at temperatures around 1.5 - 3 K, despite the observed broadening of the superconducting density of states being higher than for typical Al junctions. In addition, if the broadening could be reduced to levels commonly seen for Al, a truly wonderful device capable of reducing temperature from 1.2 K to 0.2 K would follow.   %A further reduction of  would strongly improve the performance.   %In both cases we were obliged to deposit Al on top of the superconductor due to   the difficulty associated with growing good quality respective  native oxide. Although proximity effect induced increase in the gap value of Al was observed in both case, as compared to  the simple  Cu-AlO$  _{x}$-Al  based tunnel junction grown on SiO$  _{x}$, the $r_S$ and $ \Gamma $ value of these Cu-AlO$ _{x}$-Al-NbN$ _{x}$ (Nb)   devices were much higher. In general, higher value of the $r_S$ and $ \Gamma $  degrade the device performance.  Nevertheless, we observed   while the high $r_S$  of the NbN$ _{x} $ device rendered it unfit for cooling applications \citep{NbN}. 

%Practical realization of  solid state coolers capable of reaching  0.05-0.1 K starting from a base temperature of 1.2 K (temperature that can be easily  obtained by pumping liquid He$ ^{4} $)   will simplify cryogenics tremendously. Since in such NIS devices the  peak cooling is expected to occur around 0.44 T$ _C $\citep{RMP}, materials with T$ _{C} $ $ \sim $ 3 K are  an  ideal choice for  high operational temperature coolers. If such material based devices are capable of reaching  a temperature of  $ \sim $ 0.3 K, in principle, they  can be cascaded to the already existing Al  based coolers that are capable of reaching sub  0.1 K temperatures.
  Tantalum nitride (TaN$  _{x}$)  is a  material whose T$ _C $ has been shown to be tunable in thin films between   4 - 10.8 K  by adjusting the growth parameters\citep{Kibane,JVST,Reichelt,Ilin}. Moreover, depending upon the amount of  incorporated nitrogen $ x $, TaN$  _{x}$ can be  a superconductor, insulator or a metal at low temperatures\cite{Nie,Kibane,Yu}. %The existing TaN$  _{x}$ processing  know how\cite{Shimada}  from the  micro-electronic device industry facilitates fabrication of large  junction area based NIS devices. Such large junction area devices well suited for electronic cooling and particle detection applications.  Therefore an estimate of $ r_S $ and $ \Gamma $ is essential  in order to design and obtain a rough idea of the performance of a practical device. 
In superconducting device applications, however, TaN has not been used widely. In its normal state, it has been used as a barrier material in SNS Josephson junctions with NbN \cite{Kaul,Setzu,Nevala} and NbTiN \cite{Yu2} as the superconducting electrode materials. As a superconductor, the only device application so far has been for superconducting single photon detection \cite{Engel}, and notably, no tunnel junction devices have been reported before.
%Although TaN$  _{x}$ has been widely used as a barrier material  for  tunnel junctions like  Josephson devices\cite{ Setzu,Nevala}, so far  it has not  been used as the superconductor in such devices. 
Recently,  we were able to grow high quality TaN$ _x $  thin films with $ T_{\text{C}} $ up to  $ \sim $ 8 K using a pulsed laser deposition (PLD) technique\cite{JVST}. Furthermore, we have already fabricated NbN$ _{x} $ based NIS junctions, using PLD for the growth of NbN$ _{x} $  films, and electron beam lithography (EBL), reactive ion etching (RIE), and shadow angle evaporation for the device fabrication, with an ex-situ  thermally oxidized Al barrier\citep{NbN}. These two advances were combined here to develop Cu-AlO$ _{x} $-Al-TaN$ _{x} $  NIS tunnel junctions.

\begin{figure}[htbp]
\includegraphics[width=0.9\columnwidth]{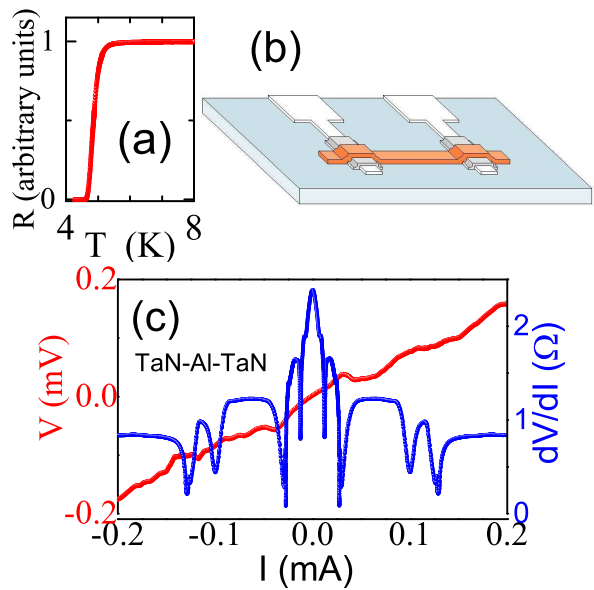}\caption{(Color online) Temperature dependence of the resistance of the bare TaN$ _{x} $ film, used for fabricating the  TaN$ _{x} $-Al-AlO$_{x}$-Cu  device, exhibiting superconducting transition at $  \sim$ 5 K.  The ac bias current used was 10 $ \mu $A. (b) A schematic of a fabricated device with two junctions. Blue: MgO substrate, White: TaN electrodes, Grey: Al/AlOx film, Orange: Cu wire. (c) Bias current dependence of the voltage and differential resistance of   a TaN$ _{x} $-Al-TaN$ _{x} $  SNS junction at 4.2 K.}\label{Fig1}
\end{figure}

First, 30 nm thick superconducting TaN$  _{x}$ films with $ T_{\text{C}} $ in the range   $ \sim $  4.5 - 5 K were deposited on  (100) oriented MgO single crystals using a PLD technique described in detail elsewhere \cite{JVST}.  A typical temperature dependence of the resistance of such a bare TaN$ _x $ film is shown in figure~\ref{Fig1}(a). MgO was chosen  as the substrate, because the films  grown on it were shown to be monophase superconducting FCC (rocksalt),  while in the films grown on  oxidized silicon  a  coexisting  non-superconducting hexagonal phase was also found \cite{JVST}.  

The TaN$ _x $ films were patterned into 1 $ \mu $m wide electrodes and large contact pads by electron beam lithography (EBL) and reactive ion etching (RIE). To make a more resistant etch mask, the TaN$ _x $ was first covered with a 50 nm  thick evaporated Cu film,  on top of which a 400 nm thick  positive PMMA resist was spun,  followed by EBL electrode patterning and removal of the Cu in the exposed regions with a chemical etch (30 \% H$_2$O$_2$, glacial acetic acid, DI water 1:1:18). After that, the exposed TaN$ _x $ was etched by RIE using CHF$  _{3}$, 50 sccm and O$  _{2}$, 5 sccm at a power 100 W and pressure 55 mTorr, the PMMA removed, and finally the remaining Cu removed by another chemical etch step.  %to obtain 1 $ \mu $m wide TaN$ _x $ electrodes and large superconducting contact pads. 

The electrode patterning was followed by the fabrication of three  distinct types of devices  using  a  second overlay EBL step and   ultra-high vacuum (UHV) e-beam evaporation. For the first device type, a 40 nm thick, 0.5 $\mu$m wide and 15 $\mu$m long aluminum cross strip was deposited across the electrodes, without any explicit attempt to form tunnel barriers. The purpose of this sample is to determine the quality of the Al-TaN$  _{x}$ contact, as an unwanted native oxide barrier may exist on the surface of the TaN$  _{x}$ film. For the second device type, the method previously developed for the NbN$ _{x} $ NIS junction fabrication with AlO$ _{x} $ tunnel barriers has been used\cite{NbN}. First, 40 nm thick Al islands of size 6 $\mu$m$\times$ 6 $\mu$m were evaporated on top the TaN$ _x $ electrodes, followed by  {\em in-situ} oxidation at room temperature in 50 mbar of O$  _{2}$  for 4 min, to grow the AlO$ _{x}$ tunnel barriers. Then, without breaking the vacuum, a 100 nm thick Cu strip of %size 15 $\mu$m$\times$ 
width 0.5 $\mu$m  was evaporated to form the connection between two TaN$ _x $ electrodes (separated by a distance of 15 $\mu$m) so that a series connection of two Cu-AlO$ _{x} $-Al-TaN$ _{x} $  NIS tunnel junctions (SINIS) is formed [Fig 1 (b)].  The third device type was identical with the second, except that no Al was deposited, and the TaN$ _{x} $ electrodes were directly oxidized in 400 mbar of O$  _{2}$  for 30 min. The goal of this process is a SINIS device  with a Cu-TaO$ _{x} $-TaN$ _{x} $ tunnel junction structure.  The typical junction dimensions  were $ \sim $ 1 $ \times $ 0.5 $ \mu $m$ ^{2} $.

Since the TaN$ _x $ films come in contact with  ambient atmosphere for prolonged periods of time during the process of fabrication, we investigated the effects of a possible native oxide barrier with the help of the first type of TaN$  _{x}$-Al-TaN$  _{x}$ device. The measured voltage and differental resistance  vs. current characteristics at 4.2 K are shown figure~\ref{Fig1}(c). Clearly, the data shows that the contact resistance is low $ <  1 \Omega$/junction (four orders of magnitude less than the tunneling resistances of the second and third type devices, as will be shown later), %Since the sample resistance was about 3 orders of magnitude smaller than the line resistance we were obliged to use the bias current as the  independent variable.   
and that the general behavior is that of a good NS contact, although several resonance features are seen, possibly originating from multiple Andreev reflections\cite{Cuevas}. The resonance features were not observed in similar devices using NbN electrodes\cite{NbN}, however, the NS contact resistance in NbN devices was actually orders of magnitude higher for unknown reasons.   %The  measured device resistance was $  \sim$ 1 $ \Omega $, which is 4 orders of magnitude less than the tunneling resistances of  Cu-AlO$ _{x}$-Al-TaN$ _{x} $ and Cu-TaO$ _{x}$-TaN$ _{x} $  based junctions, as will be seen later. 

\begin{figure}[htbp]
\includegraphics[width=0.9\columnwidth]{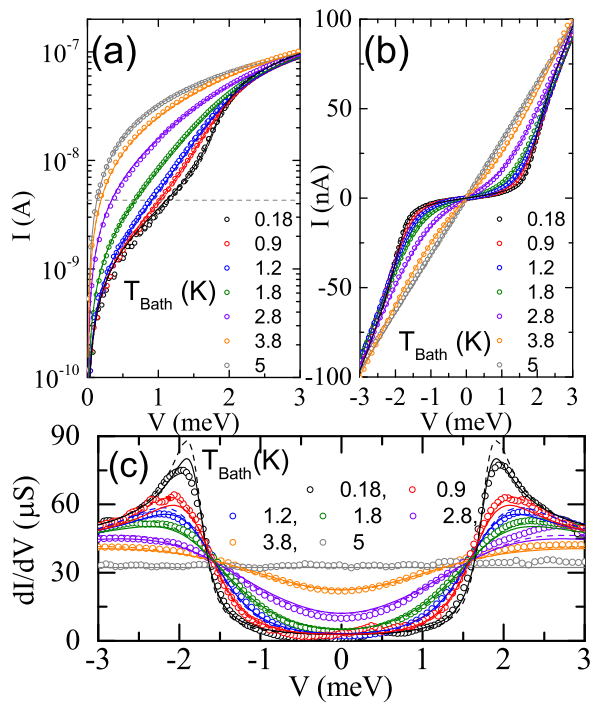}\caption{(Color online) Temperature dependence of the current-voltage characteristics of  a double junction based  Cu-AlO$ _{x}$-Al-TaN$ _{x} $ device  at various $T_{\text{Bath}}  $ plotted in (a) log-linear and (b) linear scale. (c) Differential conductance characteristics corresponding to plots in (a).    The open symbols denote the measured response, while the lines are the corresponding theoretical fits. The dashed and solid  lines corresponds to  calculations where the  tunnelling resistance of   individual junctions are assumed to be equal and unequal (proportions 66 \% and 34 \%),  respectively.  The fitting parameters were $\Gamma/\Delta(0)  $ =7$ \times $10$ ^{-2} $ and  $R_{\text{T}}$.    The horizontal dashed line in (a) represents a constant current bias of 4.3 nA. }\label{Fig2}
\end{figure}

The current-voltage and conductance-voltage measurements for the second and third type devices were carried out using a He$ ^{3} $-He$ ^{4} $ dilution refrigerator. The measurement lines had three stages of filtering: pi-filters  at 4.2 K, RC-filters at the base temperature, and microwave filtering \cite{Zorin} between these two (Thermocoax cables of length 1.5 m). For the measurement of conductance, a lock-in  technique with a 0.04 mV excitation voltage  and  17 Hz frequency   was used. 
In figure~\ref{Fig2}, the current-voltage ($I-V$) characteristics at various bath temperatures $T _{\text{Bath}} $   for a TaN$ _{x} $-Al-AlO$_{x}$-Cu based \textit{ double} junction SINIS device are shown in   (a) log-linear and (b) linear scales, respectively, together with the  corresponding theoretical fits based on the single-particle tunneling model $\frac{1}{eR_{\text{T}}}\int_{-\infty}^{\infty}\!\! d\epsilon N_{S}(\epsilon )(f_{N}(\epsilon-eV )-f_{N}(\epsilon+eV ))$,  where   $f_{N}(\epsilon)$ is the Fermi function in  Cu wire, and $N_{S}(\epsilon )$ is the normalized broadened  superconducting quasiparticle density of states (DOS)   in the Dynes  model \cite{Dynes,RMP}  $N_{S}(\epsilon, T_{S})  =\left | {\rm Re} \left ( \frac{\epsilon+i\Gamma}{\sqrt{(\epsilon+i\Gamma)^{2}-\Delta^{2}}} \right ) \right |$, where $\Gamma$ is the parameter describing broadening and $\Delta$ is the superconducting gap. The  corresponding conductance characteristics along with the theoretical fits are shown in figure~\ref{Fig2} (c).  For all these fits both the superconductor and normal metal temperatures $T_{S}$ and $ T_{N} $ were set equal to $ T_{\text{Bath}} $.  The dashed lines (most clearly visible for the lowest temperature data in (a) and (c)) assume that the tunneling resistances of the  individual junctions are identical, while the solid lines  assume non-identical tunneling resistances \citep{prb} with proportions 66 \% and 34 \% of the total resistance. This asymmetry was directly measured with the help of a third NIS junction connected to the same normal metal electrode. The individual junction resistances can then be solved from the three measurements of the SINIS pairs. The agreement between the data and the simplest possible theory with two identical barriers is already very good at higher temperatures, where the effect of the asymmetry is weaker. However, at 0.18 K, the symmetric model predicts a lower sub-gap current (visible at $V \sim 1.5$ mV) than what is observed, and the non-symmetric theory can explain this increase. It is quite important to take into account the asymmetry at low $T$: Simply fitting to the subgap conductance with $ T_{N} $ as a free parameter would give $ T_{N} \sim 0.5$ K. Such a high electronic overheating is unphysical, as it would require an excess heating power $> 100$ pW (much higher than$\sim 10 $ fW typically seen in our setup \cite{PanuPRL}), as estimated from the known electron-phonon interaction constant for Cu \cite{jenni} and the size of the normal metal island. The physical mechanism for the observed variability in $R_{\text{T}}$ is unclear, although it has been suggested \cite{greibe} that it could result from grain-to-grain barrier variability.

%The TaN$ _{x} $-Al bilayer is treated as an effective superconductor.  % This means if the total   tunneling resistances double junction  device is $ R_T $,the corresponding    individual junction resistances are   $ x R_T $ and  (1-$x R_T $).
% Clearly the asymmetric model fits the data more closely than the symmetric case. 
From the fits, we also get the temperature dependence of the superconducting gap $\Delta$, which was seen to follow the simple BCS theory well, in contrast to the NbN$ _{x} $ based devices \citep{NbN} which exhibited stronger modifications due to proximity effect \cite{Golubov}. At 0.18 K, the measured   $ \Delta$ was $ \sim $ 0.9 meV,  about four times higher than a typical Al film gap,  indicating  that the Al layer is well proximized by the TaN$ _{x} $. This value of $\Delta$ is almost the same as in the NbN NIS devices \cite{NbN} although $T_{\text{C}} = 5$ K is less than half, an observation which is consistent with the fact that the contact resistance between TaN$ _{x} $ and Al is much lower.   
%Although  proximity effects in the TaN$ _{x} $-Al bilayer is  expected to modify the shape of their  DOS  and temperature dependence of the superconducting gap, 
%we first assumed the BCS type of dependence of $ \Delta $ on  $ T_{Bath} $ and found it to fit  the observed data well. Clearly  the  temperature dependence of   $ \Delta $ of these TaN$ _{x} $ based devices   is different  from the NbN$ _{x} $ based devices we have studied earlier\citep{NbN}. This trend is consistent with theoretical predictions\cite{Golubov}  
 %as the barrier transparency of TaN$ _{x} $ devices is one order of magnitude lower than that of the NbN$ _{x} $ devices. %The $ T_C $ of this effective superconductor, used for the theoretical fits, was assumed to be 5 K - same as the  measured $ T_C $ of the TaN$ _{x} $ film.  
  All the theory fits to the $ I-V $ and conductance  curves were obtained with a broadening parameter $\Gamma/\Delta(0)$ = 7$ \times $10$^{-2}  $ [$\Delta(0)=\Delta(T=0)$], a value which is slightly higher than the smallest value observed in the NbN$ _{x} $ devices, $\Gamma_{\text{NbN}}/\Delta_{\text{NbN}} $ = 2.4$ \times $10$^{-2}  $. Similar to the NbN case, strong coupling theory did not fit the data well  (not shown). Finally,  the total  $ R_{\text{T}} $ of this device was, surprisingly, found to  evolve with temperature, from $ \sim $ 31 k$ \Omega $ at 5 K to  $ \sim $ 26.5 k$ \Omega $ at 3.8 K and $ \sim $ 24.5 k$ \Omega $ at still lower temperatures. This translates to a specific junction resistance $ r_{\text{S}} $ of  $ \sim $ 6.5  k$ \Omega\mu $m$ ^{2} $, which is about two orders of magnitude smaller than that in the NbN$ _{x} $  devices fabricated in a similar manner, but still about three-ten times higher than that of typical high power Cu-AlO$  _{x}$-Al  tunnel junction coolers \cite{Lowellnew,Nguyen}.

% To rule of any possible effect of the substrate on $ r_S $ we made several Cu-AlO$  _{x}$-Al junctions on MgO and SiN$ _{x} $ substrates where the aluminium was oxidized at 400 mbar of pure O$  _{2}$  for 20 min. The measured   $ r_S $ value of these junctions were $ \sim $ 3  k$ \Omega\mu $m$ ^{2} $ showing no noticeable influence of the substrate on $ r_S $.  

\begin{figure}[htbp]
\includegraphics[width=0.9\columnwidth]{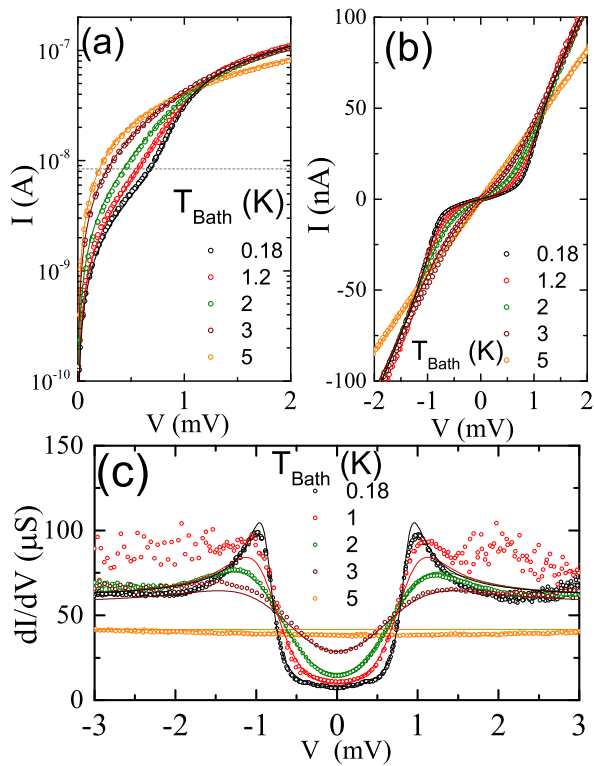}\caption{(Color online) Temperature dependence of the current-voltage characteristics of  a single junction based  Cu-TaO$ _{x}$-TaN$ _{x} $ device  at various $T_{\text{Bath}}  $ plotted in (a) log and (b) linear scale. (c) Differential conductance characteristics corresponding to plots in (a).  The dots are the experimental data while the solid lines are the corresponding theoretical fits. The fitting parameters were $\Gamma/\Delta(0)  $ =0.13 and  $R_{\text{T}}$.  The horizontal dashed line in (a) represents a constant current bias of 8.4 nA.}\label{Fig3}
\end{figure}

For the devices of the third type (TaN$ _{x} $-TaO$ _{x} $-Cu), the yield  was quite low - most of them were shorts. However, some were tunnel junctions.  In figure~\ref{Fig3} (a) and (b), the current-voltage ($I-V$) characteristics at various $T _{\text{Bath}} $ of such a  TaN$ _{x} $-TaO$ _{x} $-Cu \textit{single} NIS junction are shown in   (a) log-linear  and (b) linear scale, respectively, together with the  corresponding theoretical fits. The measured and theoretical  conductance curves are shown in figure~\ref{Fig3}(c). From the theoretical fits the obtained value of $ \Gamma/\Delta(0) $ and $ \Delta(0) $ were 0.13 and 0.87 meV respectively,  with $ T_{\text{C}} = $ 4.5 K being the  measured  value of  the TaN$ _{x} $ film.    Here, the obtained value of   $ R_{\text{T}} $  evolved with temperature even more strongly, from $ \sim $24 k$ \Omega $ at 5 K to  $ \sim $ 16.5 k$ \Omega $ below 5 K. The origin of this temperature dependence of $ R_{\text{T}}  $ is unclear to us at the moment. The largest change seems to be correlated with the transition to the superconducting state, but some temperature dependence seems to be left even at temperatures much below $T_{\text{C}}$. Although the $ \Delta $  values  of  TaN$ _x $-TaO$_{x}$-Cu and    TaN$ _x $-Al-AlO$_{x}$-Cu junctions are almost identical, the low yield and the larger value of $ \Gamma $ of the latter  render  them unfit for real device applications. 

\begin{figure}[htbp]
\includegraphics[width=0.9\columnwidth]{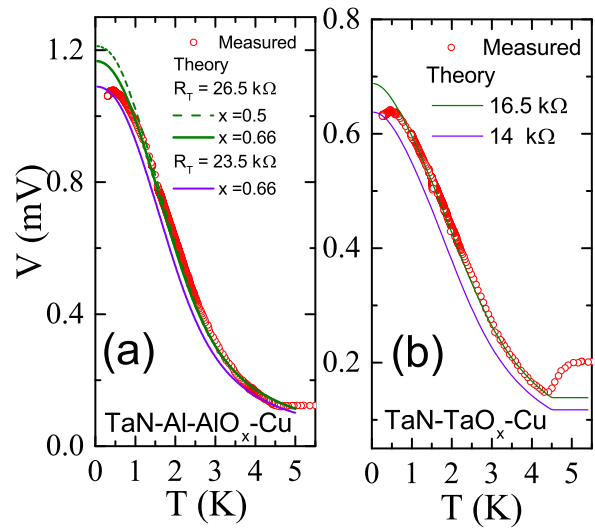}\caption{(Color online)   Thermometry characteristics of the   (a)  Cu-AlO$ _{x}$-Al-TaN$ _{x} $  junction pair biased with a constant current of $ \sim $ 4.3 nA and (b) a single  Cu-TaO$ _{x}$-TaN$ _{x} $ junction biased with  $ \sim $ 8.4 nA. The dots are the experimental data while  the  lines indicate the corresponding theoretical fits, assuming a simple one particle tunnelling model with BCS temperature dependence of the superconducting gap, for various cases of $R_{\text{T}}$. In (a),  dashed line, symmetric $R_{\text{T}} = 26.5$ k$\Omega$, solid green line, asymmetric $R_{\text{T}} = 26.5$ k$\Omega$ with with proportions 66 \% and 34 \% of the two junction resistances, solid purple line, $R_{\text{T}} = 23.5$ k$\Omega$, same asymmetry. In (b),  solid green line, $R_{\text{T}} = 16.5$ k$\Omega$, solid purple line, $R_{\text{T}} = 14$ k$\Omega$.  }\label{Fig4}
\end{figure}

Figure~\ref{Fig4} shows the thermometric response in the usual measurement configuration where the NIS junction device is current biased, and its voltage ($ V $) response is measured as a function of $ T_{\text{Bath}} $, of   the  (a) double junction TaN$ _{x} $-Al-AlO$ _{x} $-Cu device  and (b) single junction TaN$ _{x} $-TaO$ _{x} $-Cu  device. For both   devices  the measured temperature sensitivity was $ \sim $ 0.14 mV/K/junction from $ T_{\text{C}} $ down to  $ \sim $  0.5 K, as expected from theory, but at the lowest temperatures there is a saturation and even a curious downturn of the voltage. This downturn cannot be explained by any theory where $ R_{\text{T}}  $ is temperature independent for such low bias (sub-gap) values\cite{prb}, and thus the thermometry data confirms the picture of changing $R_{\text{T}}$, as can be seen from the representative theory curves. 

Having obtained the values for $ \Gamma  $  and $ r_{\text{S}} $ for the TaN NIS devices, we should compare them with previous results using other superconductors. In table~\ref{T1}, we have complied results from our lab,  fabricated in the same chamber and with fairly similar oxidation parameters. %We also list the resistivity of $\rho $ of the superconducting films. 
The parameters for TaN$ _{x} $-Al-AlO$_{x}$-Cu junctions seem to be comparable to the earlier results for Nb-Al-AlO$_{x}$-Cu junctions. The biggest difference to the standard Al-AlO$_{x}$-Cu junction technology is the much larger value of the broadening parameter $ \Gamma  $. The NbN junctions do not seem as promising for cooling as TaN junctions due to the high specific junctions resistances. The DOS broadening seen in our Al-AlO$_{x}$-Cu junctions is comparable to the results by other labs \cite{greibe,Env,oneill}. However, if extreme measures are taken to reduce environmental radiation coupling to the junction, much lower broadening has been demonstrated in Al-AlO$_{x}$-Cu junctions \cite{Env,saira}, explained by photon-assisted tunneling. According to that picture $\Gamma/\Delta \sim 1/\Delta$, which suggests that the broadening in our higher gap junctions (Nb,NbN,TaN) is due to some other mechanism.   

\begin{table}
\begin{tabular}{lrcccc}
\hline 
Junction  &$ \frac{\Gamma}{\Delta(0)} $& $ r_{\text{S}} $& $\Delta(0)$ & $T_{\text{C}}$&Ref.\tabularnewline
 &&(k$\Omega\mu$m$^{2}$)&(meV)&(K)&\tabularnewline
\hline 
\hline 
Al-AlO$_{x}$-Cu &1-5$\times  $10$ ^{-4} $& 1-2&0.21&1.4&\cite{PanuPRL}\tabularnewline
Nb-Al-AlO$_{x}$-Cu &5$\times  $10$ ^{-2} $& 10&1.0&6&\cite{Nb}\tabularnewline
NbN$ _{x} $-Al-AlO$_{x}$-Cu &2-4$\times  $10$ ^{-2} $& 630-770&1.1&10.8&\cite{NbN}\tabularnewline
NbN$ _{x} $-NbO$_{x}$-Cu &0.2& 40&1.1&10.8&\cite{NbN}\tabularnewline
TaN$ _{x} $-Al-AlO$_{x}$-Cu &7$\times  $10$^ {-2} $& 6.5&0.9&5&\tabularnewline
TaN$ _{x} $-TaO$_{x}$-Cu &0.13& 8&0.9&4.5&\tabularnewline
\hline 
\end{tabular}\caption{  Broadening of the density of states $\Gamma/\Delta(0)$,   specific tunneling  resistance $ r_{\text{S}} $, %resistivity of the superconductor $\rho $, 
energy gap at low temperature $\Delta(0)$ and critical temperature $T_{\text{C}}$ for  various types of junctions.  All these junctions were  oxidized in the same physical chamber under fairly identical oxidation conditions.}\label{T1}
\end{table}

%Having obtained  the   value of $ \Gamma $, $ r_S $, $ T_C $  and $ \Delta $ of the   TaN$ _{x} $-Al-AlO$ _{x} $-Cu junction 
 
\begin{figure}[htbp]
\includegraphics[width=0.9\columnwidth]{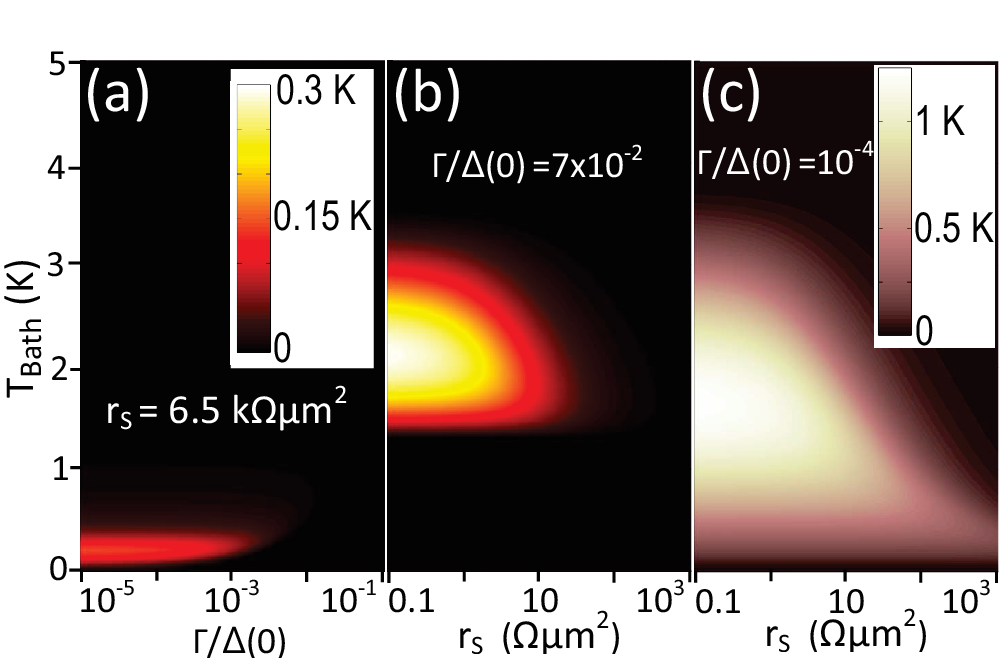}\caption{(Color online) Calculated electronic cooling ($ T_{\text{Bath}}  $ - $ T_{N}  $) for several cases, with electron-phonon limited heat transport and Cu as normal metal.  Only regions  where net cooling occur (  $ T_{\text{Bath}}  $ $ \geq $ $ T_{N}  $ ) are shown.  (a) Cooling as a function of  $ T_{\text{Bath}}  $ and $ \Gamma/\Delta(0) $  for r$ _S $ = 6.5 k$\Omega\mu$m$^{2}$. In (b) and (c), calculated cooling as a function of  $ T_{\text{Bath}}  $ and $ r_{\text{S}}$  for (b) $ \Gamma/\Delta(0)$  =7$ \times$ 10$ ^{-2} $ and   (c)  $ \Gamma/\Delta(0)$  =1$ \times$ 10$ ^{-4} $. The color bars  indicates  the magnitude of cooling. The color bar in the inset of (a) serves as the scale for both (a) and (b). It can be seen from (c) that for a device  with $ \Gamma/\Delta(0) $ = 1$ \times$ 10$ ^{-4} $  and $ r_{\text{S}} \sim$  10$\Omega\mu$m$^{2}$, the expected cooling at 1.2 K is $ \sim $ 1 K. For all simulations $T _C $ = 5 K and  $ \Delta $ = 0.9 meV. }\label{Fig5}
\end{figure}

 Finally, to answer better whether TaN based NIS junctions hold promise for cooling applications, we also carried out some numerical simulations. %to  investigate the cooling aspects of such a  device in the framework of equation~\ref{e1}.   
To give an example, all calculations assumed a SINIS device with $T_{\text{C}} = 5$ K and $\Delta = 0.9$ meV, and a Cu normal metal island of thickness 30 nm, with a lateral size the same as the total junction area $A$. Electron-phonon interaction limited heat flow out of the island was also assumed, which is the typical situation for junctions on bulk substrates \cite{R2,PanuPRL}, leading to heat balance $P_{cool}  =  \Sigma  \Omega (T_{\text{bath}}^{5}-T_{N}^5)$, where $P_{cool}$ is the cooling power of the junctions that can be calculated when junctions parameters are known \cite{R2,PanuPRL}, $\Omega$ is the normal metal volume, and $ \Sigma $ is the electron-phonon coupling constant. A typical value for $ \Sigma $ = 2$ \times $10$ ^9 $ W/(m$^{3}$K$^{5}$) in Cu was used \cite{jenni}.  %with  cross sectional area $  A$  and   overlay Cu thickness $ t $ = 30 nm ( $\Omega$ = $At$).  
Since $P_{cool} \propto A $ and  $ \Omega \propto A $, the results shown here are independent of $ A $, and therefore we use the specific junction resistance $r_{\text{S}}$ as parameter. % as long as $ A $  is small enough that Andreev current induced heating is not dominating\cite{suku,Lowell}. 
 In figure~\ref{Fig5}(a) we show the expected decrease  of $T_{N}$  below $T_{\text{Bath}}$, as a function of $ \Gamma $ and $T_{\text{Bath}}$ for the value of specific junction resistance observed in the experiment $ r_{\text{S}} $ = 6.5 k$\Omega\mu$m$^{2}$.  We find that a bit of cooling  is expected  at low $T_{\text{Bath}} \sim  0.2-0.3$ K if  $ \Gamma/\Delta(0)$ could be lowered to values  $ < 10^{-3}$. However, at that temperature range Al coolers perform better. On the other hand, if the value of $ r_{\text{S}} $ is  lowered, as shown in Fig. ~\ref{Fig5}(b), but $ \Gamma/\Delta(0)$ is fixed at the observed value 7$ \times $ 10$ ^{-2} $,  a fair amount of cooling (up to 0.3 K) at high  $T_{\text{Bath}}$ $ \sim  $ 2 - 3 K   is possible  when  $r_{\text{S}}<  $ 10$\Omega\mu$m$^{2}$. %This is shown in  figure~\ref{Fig5}(b) where the cooling is plotted as a function of  $T_{Bath}$ and $ r_S $ keeping   $ \Gamma/\Delta(0) $    fixed at 7$ \times $ 10$ ^{-2} $.
%In this case the maximum drop in $T_N $  $\sim  $0.3 K  
 Even this would fall far short from the ultimate goal to cool the metal island from 1.2 K to 0.3 K.  In order to  achieve such a large magnitude in cooling, a concomitant reduction in $ \Gamma/\Delta(0) $ of these TaN$ _{x} $ devices is also necessary. As shown if Fig. ~\ref{Fig5}(c), if  $\Gamma/\Delta(0) $ could  be lowered to    1$ \times $ 10$ ^{-4} $ (typical for Al devices), then for  $r_{\text{S}} $ $ < $  100 $\Omega\mu$m$^{2}$ such a large cooling would be theoretically possible. %this woul,  cooling the Cu island  from 1.2 K to  $ > $ 0.3 K in the framework of   equation~\ref{e1} is possible.  This can be seen from  figure~\ref{Fig5}(c) where the cooling is plotted as a function of   $T_{Bath}$ and $ r_S $ keeping the  $ \Gamma/\Delta(0) $ value  fixed at 1$ \times $ 10$ ^{-4} $. 
 Interestingly, these kind of values for $\Gamma$  and $r_{\text{S}} $ have been obtained experimentally for Al-AlO$ _{x} $-Cu  junctions. % One possible way to reduce the Dynes broadening is via improved filtering \cite{Env} and  $r_S$ can be reduced possibly by careful material consideration/growth. 

In conclusion,  we have demonstrated the application potential of normal metal-insulator-superconductor tunnel junction devices with TaN as the superconductor. The  electrical  characteristics of these devices  follow the simple one-particle  tunneling model, and the superconducting gap exhibit a BCS type temperature dependence. We also demonstrated sensitive thermometry between 0.5 and 5 K, where the lower limit was shown to be caused by an unexpected temperature variability of the tunneling resistance. The measured  effective broadening of the superconducting  density of states   and  the specific tunneling resistance  of these devices were just high enough to inhibit any electronic cooling. However, as we showed theoretically, a realistic reduction of these parameters for TaN devices would lead to a dramatic breakthrough in the development of practical electronic coolers for 1 K temperature range. Future efforts need to be especially focused on understanding the broadening of the superconducting density of states and how to reduce it.

\section*{Acknowledgements}
This research has been supported by Academy of Finland project number 260880. We thank A. Torgovkin  for help with low temperature  measurements.


%\bibliography{ref}

\begin{thebibliography}{32}%
\makeatletter
\providecommand \@ifxundefined [1]{%
 \@ifx{#1\undefined}
}%
\providecommand \@ifnum [1]{%
 \ifnum #1\expandafter \@firstoftwo
 \else \expandafter \@secondoftwo
 \fi
}%
\providecommand \@ifx [1]{%
 \ifx #1\expandafter \@firstoftwo
 \else \expandafter \@secondoftwo
 \fi
}%
\providecommand \natexlab [1]{#1}%
\providecommand \enquote  [1]{``#1''}%
\providecommand \bibnamefont  [1]{#1}%
\providecommand \bibfnamefont [1]{#1}%
\providecommand \citenamefont [1]{#1}%
\providecommand \href@noop [0]{\@secondoftwo}%
\providecommand \href [0]{\begingroup \@sanitize@url \@href}%
\providecommand \@href[1]{\@@startlink{#1}\@@href}%
\providecommand \@@href[1]{\endgroup#1\@@endlink}%
\providecommand \@sanitize@url [0]{\catcode `\\12\catcode `\$12\catcode
  `\&12\catcode `\#12\catcode `\^12\catcode `\_12\catcode `\%12\relax}%
\providecommand \@@startlink[1]{}%
\providecommand \@@endlink[0]{}%
\providecommand \url  [0]{\begingroup\@sanitize@url \@url }%
\providecommand \@url [1]{\endgroup\@href {#1}{\urlprefix }}%
\providecommand \urlprefix  [0]{URL }%
\providecommand \Eprint [0]{\href }%
\providecommand \doibase [0]{http://dx.doi.org/}%
\providecommand \selectlanguage [0]{\@gobble}%
\providecommand \bibinfo  [0]{\@secondoftwo}%
\providecommand \bibfield  [0]{\@secondoftwo}%
\providecommand \translation [1]{[#1]}%
\providecommand \BibitemOpen [0]{}%
\providecommand \bibitemStop [0]{}%
\providecommand \bibitemNoStop [0]{.\EOS\space}%
\providecommand \EOS [0]{\spacefactor3000\relax}%
\providecommand \BibitemShut  [1]{\csname bibitem#1\endcsname}%
\let\auto@bib@innerbib\@empty
%</preamble>
\bibitem [{\citenamefont {Giazotto}\ \emph {et~al.}(2006)\citenamefont
  {Giazotto}, \citenamefont {Heikkil\"a}, \citenamefont {Luukanen},
  \citenamefont {Savin},\ and\ \citenamefont {Pekola}}]{RMP}%
  \BibitemOpen
  \bibfield  {author} {\bibinfo {author} {\bibfnamefont {F.}~\bibnamefont
  {Giazotto}}, \bibinfo {author} {\bibfnamefont {T.~T.}\ \bibnamefont
  {Heikkil\"a}}, \bibinfo {author} {\bibfnamefont {A.}~\bibnamefont
  {Luukanen}}, \bibinfo {author} {\bibfnamefont {A.~M.}\ \bibnamefont {Savin}},
  \ and\ \bibinfo {author} {\bibfnamefont {J.~P.}\ \bibnamefont {Pekola}},\
  }\href {\doibase 10.1103/RevModPhys.78.217} {\bibfield  {journal} {\bibinfo
  {journal} {Rev. Mod. Phys.}\ }\textbf {\bibinfo {volume} {78}},\ \bibinfo
  {pages} {217} (\bibinfo {year} {2006})}\BibitemShut {NoStop}%
\bibitem [{\citenamefont {Muhonen}, \citenamefont {Meschke},\ and\
  \citenamefont {Pekola}(2012)}]{R2}%
  \BibitemOpen
  \bibfield  {author} {\bibinfo {author} {\bibfnamefont {J.~T.}\ \bibnamefont
  {Muhonen}}, \bibinfo {author} {\bibfnamefont {M.}~\bibnamefont {Meschke}}, \
  and\ \bibinfo {author} {\bibfnamefont {J.~P.}\ \bibnamefont {Pekola}},\
  }\href {http://stacks.iop.org/0034-4885/75/i=4/a=046501} {\bibfield
  {journal} {\bibinfo  {journal} {Rep. Prog. Phys.}\ }\textbf
  {\bibinfo {volume} {75}},\ \bibinfo {pages} {046501} (\bibinfo {year}
  {2012})}\BibitemShut {NoStop}%
\bibitem [{\citenamefont {Nahum}\ and\ \citenamefont {Martinis}(1993)}]{Nahum}%
  \BibitemOpen
  \bibfield  {author} {\bibinfo {author} {\bibfnamefont {M.}~\bibnamefont
  {Nahum}}\ and\ \bibinfo {author} {\bibfnamefont {J.~M.}\ \bibnamefont
  {Martinis}},\ }\href {\doibase 10.1063/1.110237} {\bibfield  {journal}
  {\bibinfo  {journal} {Appl. Phys. Lett.}\ }\textbf {\bibinfo {volume} {63}},\
  \bibinfo {pages} {3075} (\bibinfo {year} {1993})}\BibitemShut {NoStop}%
\bibitem [{\citenamefont {Lowell}\ \emph {et~al.}(2013)\citenamefont {Lowell},
  \citenamefont {O'Neil}, \citenamefont {Underwood},\ and\ \citenamefont
  {Ullom}}]{Lowellnew}%
  \BibitemOpen
  \bibfield  {author} {\bibinfo {author} {\bibfnamefont {P.~J.}\ \bibnamefont
  {Lowell}}, \bibinfo {author} {\bibfnamefont {G.~C.}\ \bibnamefont {O'Neil}},
  \bibinfo {author} {\bibfnamefont {J.~M.}\ \bibnamefont {Underwood}}, \ and\
  \bibinfo {author} {\bibfnamefont {J.~N.}\ \bibnamefont {Ullom}},\ }\href
  {\doibase http://dx.doi.org/10.1063/1.4793515} {\bibfield  {journal}
  {\bibinfo  {journal} {Appl. Phys. Lett.}\ }\textbf {\bibinfo {volume}
  {102}},\ \bibinfo {eid} {082601} (\bibinfo {year} {2013})}\BibitemShut
  {NoStop}%
\bibitem [{\citenamefont {Nguyen}\ \emph {et~al.}(2013)\citenamefont {Nguyen},
  \citenamefont {Aref}, \citenamefont {Kauppila}, \citenamefont {Meschke},
  \citenamefont {Winkelmann}, \citenamefont {Courtois},\ and\ \citenamefont
  {Pekola}}]{Nguyen}%
  \BibitemOpen
  \bibfield  {author} {\bibinfo {author} {\bibfnamefont {H.~Q.}\ \bibnamefont
  {Nguyen}}, \bibinfo {author} {\bibfnamefont {T.}~\bibnamefont {Aref}},
  \bibinfo {author} {\bibfnamefont {V.~J.}\ \bibnamefont {Kauppila}}, \bibinfo
  {author} {\bibfnamefont {M.}~\bibnamefont {Meschke}}, \bibinfo {author}
  {\bibfnamefont {C.~B.}\ \bibnamefont {Winkelmann}}, \bibinfo {author}
  {\bibfnamefont {H.}~\bibnamefont {Courtois}}, \ and\ \bibinfo {author}
  {\bibfnamefont {J.~P.}\ \bibnamefont {Pekola}},\ }\href@noop {} {\bibfield
  {journal} {\bibinfo  {journal} {New J. Phys.}\ }\textbf {\bibinfo {volume}
  {15}},\ \bibinfo {pages} {085013} (\bibinfo {year} {2013})}\BibitemShut
  {NoStop}%
\bibitem [{\citenamefont {Luukanen}\ \emph {et~al.}(2000)\citenamefont
  {Luukanen}, \citenamefont {Leivo}, \citenamefont {Suoknuuti}, \citenamefont
  {Manninen},\ and\ \citenamefont {Pekola}}]{Luukanen}%
  \BibitemOpen
  \bibfield  {author} {\bibinfo {author} {\bibfnamefont {A.}~\bibnamefont
  {Luukanen}}, \bibinfo {author} {\bibfnamefont {M.~M.}\ \bibnamefont {Leivo}},
  \bibinfo {author} {\bibfnamefont {J.~K.}\ \bibnamefont {Suoknuuti}}, \bibinfo
  {author} {\bibfnamefont {A.~J.}\ \bibnamefont {Manninen}}, \ and\ \bibinfo
  {author} {\bibfnamefont {J.~P.}\ \bibnamefont {Pekola}},\ }\href@noop {}
  {\bibfield  {journal} {\bibinfo  {journal} {J. Low Temp. Phys.}\ }\textbf
  {\bibinfo {volume} {120}},\ \bibinfo {pages} {281} (\bibinfo {year}
  {2000})}\BibitemShut {NoStop}%
\bibitem [{\citenamefont {Clark}\ \emph {et~al.}(2005)\citenamefont {Clark},
  \citenamefont {Miller}, \citenamefont {Williams}, \citenamefont {Ruggiero},
  \citenamefont {Hilton}, \citenamefont {Vale}, \citenamefont {Beall},
  \citenamefont {Irwin},\ and\ \citenamefont {Ullom}}]{Clark}%
  \BibitemOpen
  \bibfield  {author} {\bibinfo {author} {\bibfnamefont {A.~M.}\ \bibnamefont
  {Clark}}, \bibinfo {author} {\bibfnamefont {N.~A.}\ \bibnamefont {Miller}},
  \bibinfo {author} {\bibfnamefont {A.}~\bibnamefont {Williams}}, \bibinfo
  {author} {\bibfnamefont {S.~T.}\ \bibnamefont {Ruggiero}}, \bibinfo {author}
  {\bibfnamefont {G.~C.}\ \bibnamefont {Hilton}}, \bibinfo {author}
  {\bibfnamefont {L.~R.}\ \bibnamefont {Vale}}, \bibinfo {author}
  {\bibfnamefont {J.~A.}\ \bibnamefont {Beall}}, \bibinfo {author}
  {\bibfnamefont {K.~D.}\ \bibnamefont {Irwin}}, \ and\ \bibinfo {author}
  {\bibfnamefont {J.~N.}\ \bibnamefont {Ullom}},\ }\href {\doibase
  http://dx.doi.org/10.1063/1.1914966} {\bibfield  {journal} {\bibinfo
  {journal} {Appl. Phys. Lett.}\ }\textbf {\bibinfo {volume} {86}},\
  \bibinfo {eid} {173508} (\bibinfo {year} {2005})}\BibitemShut {NoStop}%
\bibitem [{\citenamefont {Miller}\ \emph {et~al.}(2008)\citenamefont {Miller},
  \citenamefont {O'Neil}, \citenamefont {Beall}, \citenamefont {Hilton},
  \citenamefont {Irwin}, \citenamefont {Schmidt}, \citenamefont {Vale},\ and\
  \citenamefont {Ullom}}]{Miller}%
  \BibitemOpen
  \bibfield  {author} {\bibinfo {author} {\bibfnamefont {N.~A.}\ \bibnamefont
  {Miller}}, \bibinfo {author} {\bibfnamefont {G.~C.}\ \bibnamefont {O'Neil}},
  \bibinfo {author} {\bibfnamefont {J.~A.}\ \bibnamefont {Beall}}, \bibinfo
  {author} {\bibfnamefont {G.~C.}\ \bibnamefont {Hilton}}, \bibinfo {author}
  {\bibfnamefont {K.~D.}\ \bibnamefont {Irwin}}, \bibinfo {author}
  {\bibfnamefont {D.~R.}\ \bibnamefont {Schmidt}}, \bibinfo {author}
  {\bibfnamefont {L.~R.}\ \bibnamefont {Vale}}, \ and\ \bibinfo {author}
  {\bibfnamefont {J.~N.}\ \bibnamefont {Ullom}},\ }\href {\doibase
  http://dx.doi.org/10.1063/1.2913160} {\bibfield  {journal} {\bibinfo
  {journal} {Appl. Phys. Lett.}\ }\textbf {\bibinfo {volume} {92}},\
  \bibinfo {eid} {163501} (\bibinfo {year} {2008})}\BibitemShut {NoStop}%
\bibitem [{\citenamefont {Koppinen}\ and\ \citenamefont
  {Maasilta}(2009)}]{PanuPRL}%
  \BibitemOpen
  \bibfield  {author} {\bibinfo {author} {\bibfnamefont {P.~J.}\ \bibnamefont
  {Koppinen}}\ and\ \bibinfo {author} {\bibfnamefont {I.~J.}\ \bibnamefont
  {Maasilta}},\ }\href {\doibase 10.1103/PhysRevLett.102.165502} {\bibfield
  {journal} {\bibinfo  {journal} {Phys. Rev. Lett.}\ }\textbf {\bibinfo
  {volume} {102}},\ \bibinfo {pages} {165502} (\bibinfo {year}
  {2009})}\BibitemShut {NoStop}%
\bibitem [{\citenamefont {Nevala}\ \emph {et~al.}(2012)\citenamefont {Nevala},
  \citenamefont {Chaudhuri}, \citenamefont {Halkosaari}, \citenamefont
  {Karvonen},\ and\ \citenamefont {Maasilta}}]{Nb}%
  \BibitemOpen
  \bibfield  {author} {\bibinfo {author} {\bibfnamefont {M.~R.}\ \bibnamefont
  {Nevala}}, \bibinfo {author} {\bibfnamefont {S.}~\bibnamefont {Chaudhuri}},
  \bibinfo {author} {\bibfnamefont {J.}~\bibnamefont {Halkosaari}}, \bibinfo
  {author} {\bibfnamefont {J.~T.}\ \bibnamefont {Karvonen}}, \ and\ \bibinfo
  {author} {\bibfnamefont {I.~J.}\ \bibnamefont {Maasilta}},\ }\href {\doibase
  10.1063/1.4751355} {\bibfield  {journal} {\bibinfo  {journal} {Appl.
  Phys. Lett.}\ }\textbf {\bibinfo {volume} {101}},\ \bibinfo {eid}
  {112601} (\bibinfo {year} {2012})}\BibitemShut {NoStop}%
\bibitem [{\citenamefont {Chaudhuri}, \citenamefont {Nevala},\ and\
  \citenamefont {Maasilta}(2013)}]{NbN}%
  \BibitemOpen
  \bibfield  {author} {\bibinfo {author} {\bibfnamefont {S.}~\bibnamefont
  {Chaudhuri}}, \bibinfo {author} {\bibfnamefont {M.~R.}\ \bibnamefont
  {Nevala}}, \ and\ \bibinfo {author} {\bibfnamefont {I.~J.}\ \bibnamefont
  {Maasilta}},\ }\href {\doibase http://dx.doi.org/10.1063/1.4800440}
  {\bibfield  {journal} {\bibinfo  {journal} {Appl. Phys. Lett.}\
  }\textbf {\bibinfo {volume} {102}},\ \bibinfo {eid} {132601} (\bibinfo {year}
  {2013})}\BibitemShut {NoStop}%
\bibitem [{\citenamefont {Kilbane}\ and\ \citenamefont {Habig}(1975)}]{Kibane}%
  \BibitemOpen
  \bibfield  {author} {\bibinfo {author} {\bibfnamefont {F.~M.}\ \bibnamefont
  {Kilbane}}\ and\ \bibinfo {author} {\bibfnamefont {P.~S.}\ \bibnamefont
  {Habig}},\ }\href {\doibase http://dx.doi.org/10.1116/1.568734} {\bibfield
  {journal} {\bibinfo  {journal} {J. Vac. Sci. Technol.}\
  }\textbf {\bibinfo {volume} {12}},\ \bibinfo {pages} {107} (\bibinfo {year}
  {1975})}\BibitemShut {NoStop}%
\bibitem [{\citenamefont {Chaudhuri}\ \emph {et~al.}(2013)\citenamefont
  {Chaudhuri}, \citenamefont {Maasilta}, \citenamefont {Chandernagor},
  \citenamefont {Ging},\ and\ \citenamefont {Lahtinen}}]{JVST}%
  \BibitemOpen
  \bibfield  {author} {\bibinfo {author} {\bibfnamefont {S.}~\bibnamefont
  {Chaudhuri}}, \bibinfo {author} {\bibfnamefont {I.~J.}\ \bibnamefont
  {Maasilta}}, \bibinfo {author} {\bibfnamefont {L.}~\bibnamefont
  {Chandernagor}}, \bibinfo {author} {\bibfnamefont {M.}~\bibnamefont {Ging}},
  \ and\ \bibinfo {author} {\bibfnamefont {M.}~\bibnamefont {Lahtinen}},\
  }\href {\doibase http://dx.doi.org/10.1116/1.4812698} {\bibfield  {journal}
  {\bibinfo  {journal} {J. Vac. Sci. Technol. A}\ }\textbf
  {\bibinfo {volume} {31}},\ \bibinfo {eid} {061502} (\bibinfo {year}
  {2013})}\BibitemShut {NoStop}%
\bibitem [{\citenamefont {Reichelt}, \citenamefont {Nellen},\ and\
  \citenamefont {Mair}(1978)}]{Reichelt}%
  \BibitemOpen
  \bibfield  {author} {\bibinfo {author} {\bibfnamefont {K.}~\bibnamefont
  {Reichelt}}, \bibinfo {author} {\bibfnamefont {W.}~\bibnamefont {Nellen}}, \
  and\ \bibinfo {author} {\bibfnamefont {G.}~\bibnamefont {Mair}},\ }\href
  {\doibase http://dx.doi.org/10.1063/1.324428} {\bibfield  {journal} {\bibinfo
   {journal} {J. Appl. Phys.}\ }\textbf {\bibinfo {volume} {49}},\
  \bibinfo {pages} {5284} (\bibinfo {year} {1978})}\BibitemShut {NoStop}%
\bibitem [{\citenamefont {Il'in}\ \emph {et~al.}(2012)\citenamefont
  {Il'in}, \citenamefont {Hofherr}, \citenamefont {Rall}, \citenamefont
  {Siegel}, \citenamefont {Semenov}, \citenamefont {Engel}, \citenamefont
  {Inderbitzin}, \citenamefont {Aeschbacher},\ and\ \citenamefont
  {Schilling}}]{Ilin}%
  \BibitemOpen
  \bibfield  {author} {\bibinfo {author} {\bibfnamefont {K.}~\bibnamefont
  {Il'in}}, \bibinfo {author} {\bibfnamefont {M.}~\bibnamefont {Hofherr}},
  \bibinfo {author} {\bibfnamefont {D.}~\bibnamefont {Rall}}, \bibinfo {author}
  {\bibfnamefont {M.}~\bibnamefont {Siegel}}, \bibinfo {author} {\bibfnamefont
  {A.}~\bibnamefont {Semenov}}, \bibinfo {author} {\bibfnamefont
  {A.}~\bibnamefont {Engel}}, \bibinfo {author} {\bibfnamefont
  {K.}~\bibnamefont {Inderbitzin}}, \bibinfo {author} {\bibfnamefont
  {A.}~\bibnamefont {Aeschbacher}}, \ and\ \bibinfo {author} {\bibfnamefont
  {A.}~\bibnamefont {Schilling}},\ }\href {\doibase 10.1007/s10909-011-0424-3}
  {\bibfield  {journal} {\bibinfo  {journal} {J. Low Temp.
  Phys.}\ }\textbf {\bibinfo {volume} {167}},\ \bibinfo {pages} {809}
  (\bibinfo {year} {2012})}\BibitemShut {NoStop}%
\bibitem [{\citenamefont {Nie}\ \emph {et~al.}(2001)\citenamefont {Nie},
  \citenamefont {Xu}, \citenamefont {Wang}, \citenamefont {You}, \citenamefont
  {Yang}, \citenamefont {Ong}, \citenamefont {Li},\ and\ \citenamefont
  {Liew}}]{Nie}%
  \BibitemOpen
  \bibfield  {author} {\bibinfo {author} {\bibfnamefont {H.}~\bibnamefont
  {Nie}}, \bibinfo {author} {\bibfnamefont {S.}~\bibnamefont {Xu}}, \bibinfo
  {author} {\bibfnamefont {S.}~\bibnamefont {Wang}}, \bibinfo {author}
  {\bibfnamefont {L.}~\bibnamefont {You}}, \bibinfo {author} {\bibfnamefont
  {Z.}~\bibnamefont {Yang}}, \bibinfo {author} {\bibfnamefont {C.}~\bibnamefont
  {Ong}}, \bibinfo {author} {\bibfnamefont {J.}~\bibnamefont {Li}}, \ and\
  \bibinfo {author} {\bibfnamefont {T.}~\bibnamefont {Liew}},\ }\href {\doibase
  10.1007/s003390000691} {\bibfield  {journal} {\bibinfo  {journal} {Appl.
  Phys. A}\ }\textbf {\bibinfo {volume} {73}},\ \bibinfo {pages} {229}
  (\bibinfo {year} {2001})}\BibitemShut {NoStop}%
\bibitem [{\citenamefont {Yu}\ \emph {et~al.}(2002)\citenamefont {Yu},
  \citenamefont {Stampfl}, \citenamefont {Marshall}, \citenamefont {Eshrich},
  \citenamefont {Narayanan}, \citenamefont {Rowell}, \citenamefont {Newman},\
  and\ \citenamefont {Freeman}}]{Yu}%
  \BibitemOpen
  \bibfield  {author} {\bibinfo {author} {\bibfnamefont {L.}~\bibnamefont
  {Yu}}, \bibinfo {author} {\bibfnamefont {C.}~\bibnamefont {Stampfl}},
  \bibinfo {author} {\bibfnamefont {D.}~\bibnamefont {Marshall}}, \bibinfo
  {author} {\bibfnamefont {T.}~\bibnamefont {Eshrich}}, \bibinfo {author}
  {\bibfnamefont {V.}~\bibnamefont {Narayanan}}, \bibinfo {author}
  {\bibfnamefont {J.~M.}\ \bibnamefont {Rowell}}, \bibinfo {author}
  {\bibfnamefont {N.}~\bibnamefont {Newman}}, \ and\ \bibinfo {author}
  {\bibfnamefont {A.~J.}\ \bibnamefont {Freeman}},\ }\href {\doibase
  10.1103/PhysRevB.65.245110} {\bibfield  {journal} {\bibinfo  {journal} {Phys.
  Rev. B}\ }\textbf {\bibinfo {volume} {65}},\ \bibinfo {pages} {245110}
  (\bibinfo {year} {2002})}\BibitemShut {NoStop}%
\bibitem [{\citenamefont {Kaul}\ \emph {et~al.}(2001)\citenamefont {Kaul},
  \citenamefont {Whiteley}, \citenamefont {Van~Duzer}, \citenamefont {Yu},
  \citenamefont {Newman},\ and\ \citenamefont {Rowell}}]{Kaul}%
  \BibitemOpen
  \bibfield  {author} {\bibinfo {author} {\bibfnamefont {A.~B.}\ \bibnamefont
  {Kaul}}, \bibinfo {author} {\bibfnamefont {S.~R.}\ \bibnamefont {Whiteley}},
  \bibinfo {author} {\bibfnamefont {T.}~\bibnamefont {Van~Duzer}}, \bibinfo
  {author} {\bibfnamefont {L.}~\bibnamefont {Yu}}, \bibinfo {author}
  {\bibfnamefont {N.}~\bibnamefont {Newman}}, \ and\ \bibinfo {author}
  {\bibfnamefont {J.~M.}\ \bibnamefont {Rowell}},\ }\href {\doibase
  http://dx.doi.org/10.1063/1.1337630} {\bibfield  {journal} {\bibinfo
  {journal} {Appl. Phys. Lett.}\ }\textbf {\bibinfo {volume} {78}},\
  \bibinfo {pages} {99} (\bibinfo {year} {2001})}\BibitemShut {NoStop}%
\bibitem [{\citenamefont {Setzu}, \citenamefont {Baggetta},\ and\ \citenamefont
  {Villegier}(2008)}]{Setzu}%
  \BibitemOpen
  \bibfield  {author} {\bibinfo {author} {\bibfnamefont {R.}~\bibnamefont
  {Setzu}}, \bibinfo {author} {\bibfnamefont {E.}~\bibnamefont {Baggetta}}, \
  and\ \bibinfo {author} {\bibfnamefont {J.~C.}\ \bibnamefont {Villegier}},\
  }\href {http://stacks.iop.org/1742-6596/97/i=1/a=012077} {\bibfield
  {journal} {\bibinfo  {journal} {J. Phys: Conf. Series}\
  }\textbf {\bibinfo {volume} {97}},\ \bibinfo {pages} {012077} (\bibinfo
  {year} {2008})}\BibitemShut {NoStop}%
\bibitem [{\citenamefont {Nevala}\ \emph {et~al.}(2009)\citenamefont {Nevala},
  \citenamefont {Maasilta}, \citenamefont {Senapati},\ and\ \citenamefont
  {Budhani}}]{Nevala}%
  \BibitemOpen
  \bibfield  {author} {\bibinfo {author} {\bibfnamefont {M.}~\bibnamefont
  {Nevala}}, \bibinfo {author} {\bibfnamefont {I.}~\bibnamefont {Maasilta}},
  \bibinfo {author} {\bibfnamefont {K.}~\bibnamefont {Senapati}}, \ and\
  \bibinfo {author} {\bibfnamefont {R.}~\bibnamefont {Budhani}},\ }\href
  {\doibase 10.1109/TASC.2009.2019030} {\bibfield  {journal} {\bibinfo
  {journal} { IEEE Trans. Appl. Supercond.}\ }\textbf
  {\bibinfo {volume} {19}},\ \bibinfo {pages} {253} (\bibinfo {year}
  {2009})}\BibitemShut {NoStop}%
\bibitem [{\citenamefont {Yu}\ \emph {et~al.}(2006)\citenamefont {Yu},
  \citenamefont {Gandikota}, \citenamefont {Singh}, \citenamefont {Gu},
  \citenamefont {Smith}, \citenamefont {Meng}, \citenamefont {Zeng},
  \citenamefont {Rowell},\ and\ \citenamefont {Newman}}]{Yu2}%
  \BibitemOpen
  \bibfield  {author} {\bibinfo {author} {\bibfnamefont {L.}~\bibnamefont
  {Yu}}, \bibinfo {author} {\bibfnamefont {R.}~\bibnamefont {Gandikota}},
  \bibinfo {author} {\bibfnamefont {R.~K.}\ \bibnamefont {Singh}}, \bibinfo
  {author} {\bibfnamefont {L.}~\bibnamefont {Gu}}, \bibinfo {author}
  {\bibfnamefont {D.~J.}\ \bibnamefont {Smith}}, \bibinfo {author}
  {\bibfnamefont {X.}~\bibnamefont {Meng}}, \bibinfo {author} {\bibfnamefont
  {T.}~\bibnamefont {Zeng}, \bibfnamefont {X.~Van~Duzer}}, \bibinfo {author}
  {\bibfnamefont {J.~M.}\ \bibnamefont {Rowell}}, \ and\ \bibinfo {author}
  {\bibfnamefont {N.}~\bibnamefont {Newman}},\ }\href@noop {} {\bibfield
  {journal} {\bibinfo  {journal} {Supercond. Sci. Technol.}\ }\textbf {\bibinfo
  {volume} {19}},\ \bibinfo {pages} {719} (\bibinfo {year} {2006})}\BibitemShut
  {NoStop}%
\bibitem [{\citenamefont {Engel}\ \emph {et~al.}(2012)\citenamefont {Engel},
  \citenamefont {Aeschbacher}, \citenamefont {Inderbitzin}, \citenamefont
  {Schilling}, \citenamefont {Il'in}, \citenamefont {Hofherr}, \citenamefont
  {Siegel}, \citenamefont {Semenov},\ and\ \citenamefont {H\"ubers}}]{Engel}%
  \BibitemOpen
  \bibfield  {author} {\bibinfo {author} {\bibfnamefont {A.}~\bibnamefont
  {Engel}}, \bibinfo {author} {\bibfnamefont {A.}~\bibnamefont {Aeschbacher}},
  \bibinfo {author} {\bibfnamefont {K.}~\bibnamefont {Inderbitzin}}, \bibinfo
  {author} {\bibfnamefont {A.}~\bibnamefont {Schilling}}, \bibinfo {author}
  {\bibfnamefont {K.}~\bibnamefont {Il'in}}, \bibinfo {author} {\bibfnamefont
  {M.}~\bibnamefont {Hofherr}}, \bibinfo {author} {\bibfnamefont
  {M.}~\bibnamefont {Siegel}}, \bibinfo {author} {\bibfnamefont
  {A.}~\bibnamefont {Semenov}}, \ and\ \bibinfo {author} {\bibfnamefont
  {H.-W.}\ \bibnamefont {H\"ubers}},\ }\href {\doibase
  http://dx.doi.org/10.1063/1.3684243} {\bibfield  {journal} {\bibinfo
  {journal} {Appl. Phys. Lett.}\ }\textbf {\bibinfo {volume} {100}},\
  \bibinfo {eid} {062601} (\bibinfo {year} {2012})}\BibitemShut {NoStop}%
\bibitem [{\citenamefont {Cuevas}\ \emph {et~al.}(2006)\citenamefont {Cuevas},
  \citenamefont {Hammer}, \citenamefont {Kopu}, \citenamefont {Viljas},\ and\
  \citenamefont {Eschrig}}]{Cuevas}%
  \BibitemOpen
  \bibfield  {author} {\bibinfo {author} {\bibfnamefont {J.~C.}\ \bibnamefont
  {Cuevas}}, \bibinfo {author} {\bibfnamefont {J.}~\bibnamefont {Hammer}},
  \bibinfo {author} {\bibfnamefont {J.}~\bibnamefont {Kopu}}, \bibinfo {author}
  {\bibfnamefont {J.~K.}\ \bibnamefont {Viljas}}, \ and\ \bibinfo {author}
  {\bibfnamefont {M.}~\bibnamefont {Eschrig}},\ }\href {\doibase
  10.1103/PhysRevB.73.184505} {\bibfield  {journal} {\bibinfo  {journal} {Phys.
  Rev. B}\ }\textbf {\bibinfo {volume} {73}},\ \bibinfo {pages} {184505}
  (\bibinfo {year} {2006})}\BibitemShut {NoStop}%
\bibitem [{\citenamefont {Zorin}(1995)}]{Zorin}%
  \BibitemOpen
  \bibfield  {author} {\bibinfo {author} {\bibfnamefont {A.~B.}\ \bibnamefont
  {Zorin}},\ }\href {\doibase 10.1063/1.1145385} {\bibfield  {journal}
  {\bibinfo  {journal} {Rev. Sci. Instrum.}\ }\textbf {\bibinfo
  {volume} {66}},\ \bibinfo {pages} {4296} (\bibinfo {year}
  {1995})}\BibitemShut {NoStop}%
\bibitem [{\citenamefont {Dynes}\ \emph {et~al.}(1984)\citenamefont {Dynes},
  \citenamefont {Garno}, \citenamefont {Hertel},\ and\ \citenamefont
  {Orlando}}]{Dynes}%
  \BibitemOpen
  \bibfield  {author} {\bibinfo {author} {\bibfnamefont {R.~C.}\ \bibnamefont
  {Dynes}}, \bibinfo {author} {\bibfnamefont {J.~P.}\ \bibnamefont {Garno}},
  \bibinfo {author} {\bibfnamefont {G.~B.}\ \bibnamefont {Hertel}}, \ and\
  \bibinfo {author} {\bibfnamefont {T.~P.}\ \bibnamefont {Orlando}},\ }\href
  {\doibase 10.1103/PhysRevLett.53.2437} {\bibfield  {journal} {\bibinfo
  {journal} {Phys. Rev. Lett.}\ }\textbf {\bibinfo {volume} {53}},\ \bibinfo
  {pages} {2437} (\bibinfo {year} {1984})}\BibitemShut {NoStop}%
\bibitem [{\citenamefont {Chaudhuri}\ and\ \citenamefont
  {Maasilta}(2012)}]{prb}%
  \BibitemOpen
  \bibfield  {author} {\bibinfo {author} {\bibfnamefont {S.}~\bibnamefont
  {Chaudhuri}}\ and\ \bibinfo {author} {\bibfnamefont {I.~J.}\ \bibnamefont
  {Maasilta}},\ }\href {\doibase 10.1103/PhysRevB.85.014519} {\bibfield
  {journal} {\bibinfo  {journal} {Phys. Rev. B}\ }\textbf {\bibinfo {volume}
  {85}},\ \bibinfo {pages} {014519} (\bibinfo {year} {2012})}\BibitemShut
  {NoStop}%
\bibitem [{\citenamefont {Karvonen}, \citenamefont {Taskinen},\ and\
  \citenamefont {Maasilta}(2007)}]{jenni}%
  \BibitemOpen
  \bibfield  {author} {\bibinfo {author} {\bibfnamefont {J.~T.}\ \bibnamefont
  {Karvonen}}, \bibinfo {author} {\bibfnamefont {L.~J.}\ \bibnamefont
  {Taskinen}}, \ and\ \bibinfo {author} {\bibfnamefont {I.~J.}\ \bibnamefont
  {Maasilta}},\ }\href@noop {} {\bibfield  {journal} {\bibinfo  {journal} {J.
  Low Temp. Phys.}\ }\textbf {\bibinfo {volume} {149}},\ \bibinfo {pages} {121}
  (\bibinfo {year} {2007})}\BibitemShut {NoStop}%
\bibitem [{\citenamefont {Greibe}\ \emph {et~al.}(2011)\citenamefont {Greibe},
  \citenamefont {Stenberg}, \citenamefont {Wilson}, \citenamefont {Bauch},
  \citenamefont {Shumeiko},\ and\ \citenamefont {Delsing}}]{greibe}%
  \BibitemOpen
  \bibfield  {author} {\bibinfo {author} {\bibfnamefont {T.}~\bibnamefont
  {Greibe}}, \bibinfo {author} {\bibfnamefont {M.~P.~V.}\ \bibnamefont
  {Stenberg}}, \bibinfo {author} {\bibfnamefont {C.~M.}\ \bibnamefont
  {Wilson}}, \bibinfo {author} {\bibfnamefont {T.}~\bibnamefont {Bauch}},
  \bibinfo {author} {\bibfnamefont {V.~S.}\ \bibnamefont {Shumeiko}}, \ and\
  \bibinfo {author} {\bibfnamefont {P.}~\bibnamefont {Delsing}},\ }\href@noop
  {} {\bibfield  {journal} {\bibinfo  {journal} {Phys. Rev. Lett.}\ }\textbf
  {\bibinfo {volume} {106}},\ \bibinfo {pages} {097001} (\bibinfo {year}
  {2011})}\BibitemShut {NoStop}%
\bibitem [{\citenamefont {Golubov}\ \emph {et~al.}(1995)\citenamefont
  {Golubov}, \citenamefont {Houwman}, \citenamefont {Gijsbertsen},
  \citenamefont {Krasnov}, \citenamefont {Flokstra}, \citenamefont {Rogalla},\
  and\ \citenamefont {Kupriyanov}}]{Golubov}%
  \BibitemOpen
  \bibfield  {author} {\bibinfo {author} {\bibfnamefont {A.~A.}\ \bibnamefont
  {Golubov}}, \bibinfo {author} {\bibfnamefont {E.~P.}\ \bibnamefont
  {Houwman}}, \bibinfo {author} {\bibfnamefont {J.~G.}\ \bibnamefont
  {Gijsbertsen}}, \bibinfo {author} {\bibfnamefont {V.~M.}\ \bibnamefont
  {Krasnov}}, \bibinfo {author} {\bibfnamefont {J.}~\bibnamefont {Flokstra}},
  \bibinfo {author} {\bibfnamefont {H.}~\bibnamefont {Rogalla}}, \ and\
  \bibinfo {author} {\bibfnamefont {M.~Y.}\ \bibnamefont {Kupriyanov}},\ }\href
  {\doibase 10.1103/PhysRevB.51.1073} {\bibfield  {journal} {\bibinfo
  {journal} {Phys. Rev. B}\ }\textbf {\bibinfo {volume} {51}},\ \bibinfo
  {pages} {1073} (\bibinfo {year} {1995})}\BibitemShut {NoStop}%
\bibitem [{\citenamefont {Pekola}\ \emph {et~al.}(2010)\citenamefont {Pekola},
  \citenamefont {Maisi}, \citenamefont {Kafanov}, \citenamefont {Chekurov},
  \citenamefont {Kemppinen}, \citenamefont {Pashkin}, \citenamefont {Saira},
  \citenamefont {M\"ott\"onen},\ and\ \citenamefont {Tsai}}]{Env}%
  \BibitemOpen
  \bibfield  {author} {\bibinfo {author} {\bibfnamefont {J.~P.}\ \bibnamefont
  {Pekola}}, \bibinfo {author} {\bibfnamefont {V.~F.}\ \bibnamefont {Maisi}},
  \bibinfo {author} {\bibfnamefont {S.}~\bibnamefont {Kafanov}}, \bibinfo
  {author} {\bibfnamefont {N.}~\bibnamefont {Chekurov}}, \bibinfo {author}
  {\bibfnamefont {A.}~\bibnamefont {Kemppinen}}, \bibinfo {author}
  {\bibfnamefont {Y.~A.}\ \bibnamefont {Pashkin}}, \bibinfo {author}
  {\bibfnamefont {O.-P.}\ \bibnamefont {Saira}}, \bibinfo {author}
  {\bibfnamefont {M.}~\bibnamefont {M\"ott\"onen}}, \ and\ \bibinfo {author}
  {\bibfnamefont {J.~S.}\ \bibnamefont {Tsai}},\ }\href {\doibase
  10.1103/PhysRevLett.105.026803} {\bibfield  {journal} {\bibinfo  {journal}
  {Phys. Rev. Lett.}\ }\textbf {\bibinfo {volume} {105}},\ \bibinfo {pages}
  {026803} (\bibinfo {year} {2010})}\BibitemShut {NoStop}%
\bibitem [{\citenamefont {O'Neil}\ \emph {et~al.}(2012)\citenamefont {O'Neil},
  \citenamefont {Lowell}, \citenamefont {Underwood},\ and\ \citenamefont
  {Ullom}}]{oneill}%
  \BibitemOpen
  \bibfield  {author} {\bibinfo {author} {\bibfnamefont {G.~C.}\ \bibnamefont
  {O'Neil}}, \bibinfo {author} {\bibfnamefont {P.~J.}\ \bibnamefont {Lowell}},
  \bibinfo {author} {\bibfnamefont {J.~M.}\ \bibnamefont {Underwood}}, \ and\
  \bibinfo {author} {\bibfnamefont {J.~N.}\ \bibnamefont {Ullom}},\ }\href@noop
  {} {\bibfield  {journal} {\bibinfo  {journal} {Phys. Rev. B}\ }\textbf
  {\bibinfo {volume} {85}},\ \bibinfo {pages} {134504} (\bibinfo {year}
  {2012})}\BibitemShut {NoStop}%
\bibitem [{\citenamefont {Saira}\ \emph {et~al.}(2012)\citenamefont {Saira},
  \citenamefont {Kemppinen}, \citenamefont {Maisi},\ and\ \citenamefont
  {Pekola}}]{saira}%
  \BibitemOpen
  \bibfield  {author} {\bibinfo {author} {\bibfnamefont {O.-P.}\ \bibnamefont
  {Saira}}, \bibinfo {author} {\bibfnamefont {A.}~\bibnamefont {Kemppinen}},
  \bibinfo {author} {\bibfnamefont {V.~F.}\ \bibnamefont {Maisi}}, \ and\
  \bibinfo {author} {\bibfnamefont {J.~P.}\ \bibnamefont {Pekola}},\
  }\href@noop {} {\bibfield  {journal} {\bibinfo  {journal} {Phys. Rev. B}\
  }\textbf {\bibinfo {volume} {85}},\ \bibinfo {pages} {012504} (\bibinfo
  {year} {2012})}\BibitemShut {NoStop}%
\end{thebibliography}

\newpage

\end{document}